\documentclass[twocolumn]{aastex62}

\usepackage{graphicx}
\usepackage{txfonts}

\usepackage[capitalise]{cleveref}

\def\kms{\,km\,s$^{-1}$}       
\def\msol{$M_\odot$}		
\def\rsol{$R_\odot$}		
\def\mstar{$M_*$}		
\def\rstar{$R_*$}		
\def\densstar{$\rho_*$}		
\def\mplanet{$M_{\rm P}$}	
\def\mjup{$M_{\rm Jup}$}	
\def\teql{$T_{\rm eql}$}
\def\teff{$T_{\rm eff}$}
\def\feh{[Fe/H]}
\def\logg{$\log g_*$}

\def\kms{km\, s$^{-1}$}
\def\sun{$\odot$}

\def\ecos{$e \cos \omega$}
\def\esin{$e \sin \omega$}
\def\secos{$\sqrt{e} \cos \omega$}
\def\sesin{$\sqrt{e} \sin \omega$}
\def\vsicos{$v_* \sin i_* \cos \lambda$}
\def\vsisin{$v_* \sin i_* \sin \lambda$}
\def\svsicos{$\sqrt{v_* \sin i_*} \cos \lambda$}
\def\svsisin{$\sqrt{v_* \sin i_*} \sin \lambda$}

\shorttitle{A low-density hot Jupiter in a near-aligned orbit}
\shortauthors{Anderson et al.}

\begin{document}

\title{The discovery of WASP-134b, WASP-134c, WASP-137b, WASP-143b and WASP-146b: three hot Jupiters and a pair of warm Jupiters orbiting Solar-type stars}

\correspondingauthor{D. R. Anderson}
\email{d.r.anderson@keele.ac.uk}

\author[0000-0001-7416-7522]{D. R. Anderson}
\affiliation{Astrophysics Group, Keele University, Staffordshire ST5 5BG, UK}

\author{F. Bouchy}
\affiliation{Observatoire de Gen\`eve, Universit\'e de Gen\`eve, 51 Chemin 
       des Maillettes, 1290 Sauverny, Switzerland}

\author{D. J. A. Brown}
\affiliation{Department of Physics, University of Warwick, Coventry CV4 7AL, UK}
\affiliation{Centre for Exoplanets and Habitability, University of Warwick, Gibbet Hill Road, Coventry CV4 7AL, UK}

\author{A.~Collier~Cameron}
\affiliation{SUPA, School of Physics and Astronomy, University of St. Andrews, 
       North Haugh, Fife KY16 9SS, UK}

\author{L.~Delrez}
\affiliation{Space sciences, Technologies and Astrophysics Research (STAR) Institute, Universit\'e de Li\`ege, Li\`ege 1, Belgium}
\affiliation{Cavendish Laboratory, J J Thomson Avenue, Cambridge CB3 0HE, UK}

\author{M.~Gillon}
\affiliation{Space sciences, Technologies and Astrophysics Research (STAR) Institute, Universit\'e de Li\`ege, Li\`ege 1, Belgium}

\author{J. I. Gonz\'alez Hern\'andez}
\affiliation{Instituto de Astrof\'isica de Canarias, E-38205 La Laguna, Tenerife, Spain}
\affiliation{Universidad de La Laguna, Departamento de Astrof\'isica, E-38206 La Laguna, Tenerife, Spain}

\author{C.~Hellier}
\affiliation{Astrophysics Group, Keele University, Staffordshire ST5 5BG, UK}

\author{E.~Jehin}
\affiliation{Space sciences, Technologies and Astrophysics Research (STAR) Institute, Universit\'e de Li\`ege, Li\`ege 1, Belgium}

\author{M.~Lendl}
\affiliation{Observatoire de Gen\`eve, Universit\'e de Gen\`eve, 51 Chemin 
       des Maillettes, 1290 Sauverny, Switzerland}
\affiliation{Space Research Institute, Austrian Academy of Sciences, Schmiedlstr. 6, 8042 Graz, Austria}

\author{P.~F.~L.~Maxted}
\affiliation{Astrophysics Group, Keele University, Staffordshire ST5 5BG, UK}

\author{M.~Neveu-VanMalle}
\affiliation{Observatoire de Gen\`eve, Universit\'e de Gen\`eve, 51 Chemin 
       des Maillettes, 1290 Sauverny, Switzerland}

\author{L.~D.~Nielsen}
\affiliation{Observatoire de Gen\`eve, Universit\'e de Gen\`eve, 51 Chemin 
       des Maillettes, 1290 Sauverny, Switzerland}

\author{F.~Pepe}
\affiliation{Observatoire de Gen\`eve, Universit\'e de Gen\`eve, 51 Chemin 
       des Maillettes, 1290 Sauverny, Switzerland}

\author{M. Perger}
\affiliation{Institut de Ci\`encies de l'Espai, Campus UAB, C/Can Magrans s/n, 08193 Bellaterra, Spain}
\affiliation{Institut d'Estudis Espacials de Catalunya (IEEC), 08034 Barcelona, Spain}

\author{D.~Pollacco}
\affiliation{Department of Physics, University of Warwick, Coventry CV4 7AL, UK}
\affiliation{Centre for Exoplanets and Habitability, University of Warwick, Gibbet Hill Road, Coventry CV4 7AL, UK}

\author{D.~Queloz}
\affiliation{Observatoire de Gen\`eve, Universit\'e de Gen\`eve, 51 Chemin 
       des Maillettes, 1290 Sauverny, Switzerland}
\affiliation{Cavendish Laboratory, J J Thomson Avenue, Cambridge CB3 0HE, UK}

\author{J.~Rey}
\affiliation{Observatoire de Gen\`eve, Universit\'e de Gen\`eve, 51 Chemin 
       des Maillettes, 1290 Sauverny, Switzerland}

\author{D.~S\'egransan}
\affiliation{Observatoire de Gen\`eve, Universit\'e de Gen\`eve, 51 Chemin 
       des Maillettes, 1290 Sauverny, Switzerland}

\author{B.~Smalley}
\affiliation{Astrophysics Group, Keele University, Staffordshire ST5 5BG, UK}

\author{B.~Toledo-Padr\'on}
\affiliation{Instituto de Astrof\'isica de Canarias, E-38205 La Laguna, Tenerife, Spain}
\affiliation{Universidad de La Laguna, Departamento de Astrof\'isica, E-38206 La Laguna, Tenerife, Spain}

\author{A.~H.~M.~J.~Triaud}
\affiliation{School of Physics \& Astronomy, University of Birmingham, Edgbaston, Birmingham, B15 2TT, UK}

\author{O. D. Turner}
\affiliation{Observatoire de Gen\`eve, Universit\'e de Gen\`eve, 51 Chemin 
       des Maillettes, 1290 Sauverny, Switzerland}

\author{S.~Udry}
\affiliation{Observatoire de Gen\`eve, Universit\'e de Gen\`eve, 51 Chemin 
       des Maillettes, 1290 Sauverny, Switzerland}

\author{R.~G.~West}
\affiliation{Department of Physics, University of Warwick, Coventry CV4 7AL, UK}
\affiliation{Centre for Exoplanets and Habitability, University of Warwick, Gibbet Hill Road, Coventry CV4 7AL, UK}



\begin{abstract}
We report the discovery by WASP of five planets orbiting moderately bright ($V$ = 11.0--12.9) Solar-type stars. WASP-137b, WASP-143b and WASP-146b are typical hot Jupiters in orbits of 3--4\,d and with masses in the range 0.68--1.11\,\mjup. 
WASP-134 is a metal-rich ([Fe/H] = +0.40 $\pm$ 0.07]) G4 star orbited by two warm Jupiters: WASP-134b (\mplanet\ = 1.41 \mjup; $P = 10.1$\,d; $e = 0.15 \pm 0.01$; \teql\ = 950\,K) and WASP-134c (\mplanet\ $\sin i$ = 0.70 \mjup; $P = 70.0$\,d; $e = 0.17 \pm 0.09$; \teql\ = 500\,K).
From observations of the Rossiter-McLaughlin effect of WASP-134b, we find its orbit to be misaligned with the spin of its star ($\lambda = -44 \pm 10^\circ$). 
WASP-134 is a rare example of a system with a short-period giant planet and a nearby giant companion. In-situ formation or disc migration seem more likely explanations for such systems than does high-eccentricity migration.
\end{abstract}

\keywords{planets and satellites: individual (WASP-134b, WASP-134c, WASP-137b, WASP-143b, WASP-146b)}

\section{Introduction} \label{sec:intro}
As we near a tally of 200 planets discovered by our ground-based transit survey WASP \citep{2006PASP..118.1407P}, TESS is ushering in the era of space-based wide-field transit surveys \citep{2014SPIE.9143E..20R}. 
TESS will provide an important test of the completeness of WASP, as well as of similar surveys such as KELT \citep{2007PASP..119..923P}, HATNet and HATSouth \citep{2018haex.bookE.111B}.

Considering that during its nominal mission TESS will observe most of its target fields for only 27 days, it faces a challenge to discover planets with periods longer than half that duration. 
For some of those planets for which TESS observes only one or two transits, ephemerides may be recovered by combining the TESS data with the long baseline of the aforementioned surveys \citep{2018arXiv180711922Y}. 
Though each of those surveys is multi-site, even single-site observations would be effective in following up TESS's single-transit detections \citep{2018A&A...619A.175C}. 
Thus ground-based projects such as TRAPPIST \citep{2011EPJWC..1106002G,2011Msngr.145....2J}, NGTS \citep{2018MNRAS.475.4476W} and SPECULOOS \citep{2018SPIE10700E..1ID} stand to play an important role, as does ESA's CHEOPS satellite \citep{2013EPJWC..4703005B}, which is scheduled for launch in late 2019.

Warm Jupiters (orbital period $P$ = 10--200\,d) are more difficult to find than are hot Jupiters, due to their lower geometric transit probability, less frequent transits, and longer transit durations. Also, they may be inherently less common (e.g. \citealt{2016A&A...587A..64S}). Considering only `Jupiters' (planets with mass \mplanet\ $>$ 0.3\,\mjup), the TEPCat database lists 416 hot Jupiters ($P < 10$\,d), but only 37 Jupiters  in the range $P$ = 10--20\,d \citep{2011MNRAS.417.2166S}. 

Longer-period planets are certainly worth the effort required to find them: to gain an understanding of the diversity of exoplanets, their various formation and evolution histories, bulk and atmospheric compositions, etc., we must populate a wide region of parameter space. 
Specifically, by measuring the orbital eccentricities and stellar obliquities of planets in the tail end of the hot-Jupiter period distribution, where tides are weak, it may be that we can explain hot Jupiter migration (e.g. \citealt{2015ApJ...800L...9A}).

While it is common for a hot Jupiter to have a massive companion in a wide orbit (e.g. \citealt{2012ApJ...749..134H,2016A&A...586A..93N,2017MNRAS.467.1714T}), no hot Jupiters are known to have close companions (with the notable exception of WASP-47b; \citealt{2012MNRAS.426..739H,2015ApJ...812L..18B}). Conversely, half of warm Jupiters are flanked by small companions, which \citet{2016ApJ...825...98H} interpreted as indicating that warm Jupiters formed in situ. 

We report here the discovery by the WASP survey of five planets orbiting Sun-like stars. In three systems we detect a single hot Jupiter and in the fourth system we detect two warm Jupiters. We measure the orbital eccentricity of both warm Jupiters and the stellar obliquity of the inner planet.

\section{Observations} \label{sec:obs}
\begin{figure}
\includegraphics[width=84mm]{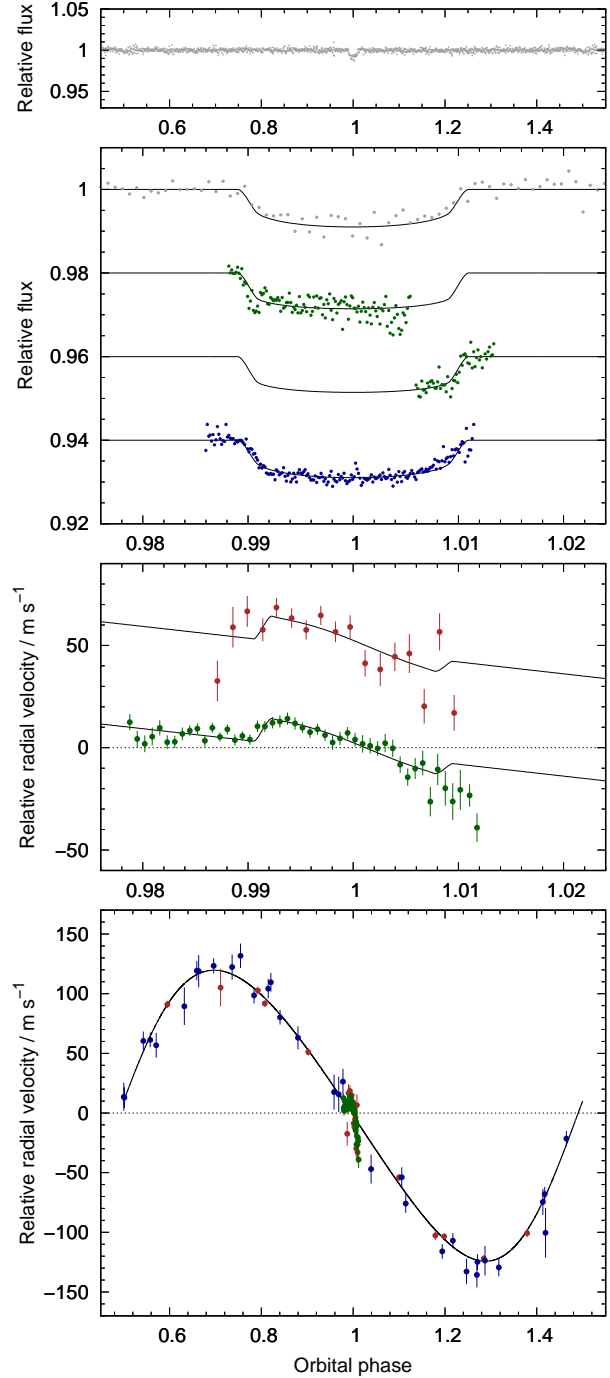}
\caption{WASP-134b discovery data. 
{\it Top panel}: WASP lightcurve folded on the transit ephemeris and binned with a bin width of 10\,min. 
{\it Second panel}: Transit lightcurves from WASP (grey), TRAPPIST-South (green) and EulerCam (blue), offset for clarity, binned with a bin width of 2\,min (10\,min for WASP), and plotted chronologically with the most recent lightcurve at the bottom. 
The best-fitting transit model is superimposed. 
{\it Third panel}: The RVs from HARPS-S (brown) and HARPS-N (green) showing the apparent anomaly during transit, along with the best-fitting Rossiter-McLaughlin (RM) effect model.
{\it Bottom panel}: The RVs from CORALIE (blue), HARPS-S and HARPS-N, with the best-fitting eccentric orbital model. 
WASP-134 reached high airmass by the end of each photometric and spectroscopic transit sequence. 
\label{fig:w134-rv-phot}}
\end{figure}

\begin{figure}
\includegraphics[width=84mm]{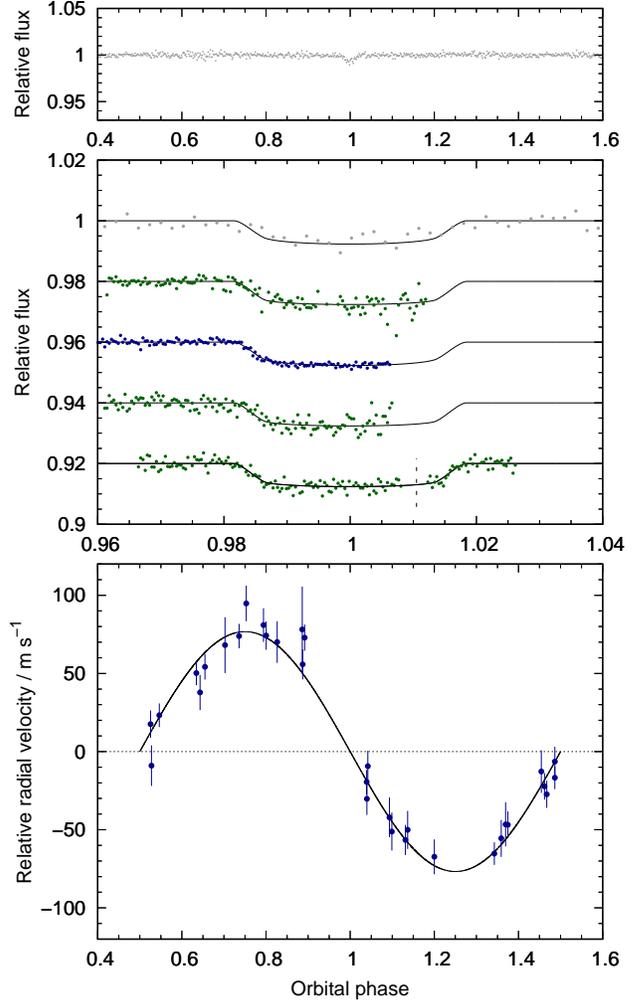}
\caption{WASP-137b discovery data. 
As for \cref{fig:w134-rv-phot}.
The meridian flip is indicated with a vertical dashed line.\label{fig:w137-rv-phot}}
\end{figure}

\begin{figure}
\includegraphics[width=84mm]{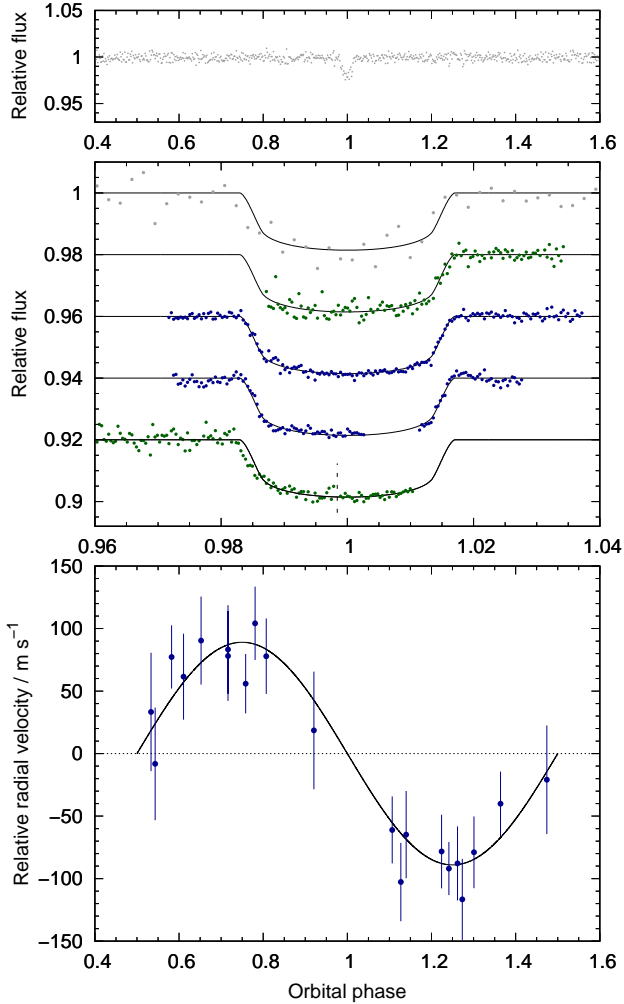}
\caption{WASP-143b discovery data. 
As for \cref{fig:w137-rv-phot}.
\label{fig:w143-rv-phot}}
\end{figure}

\begin{figure}
\includegraphics[width=84mm]{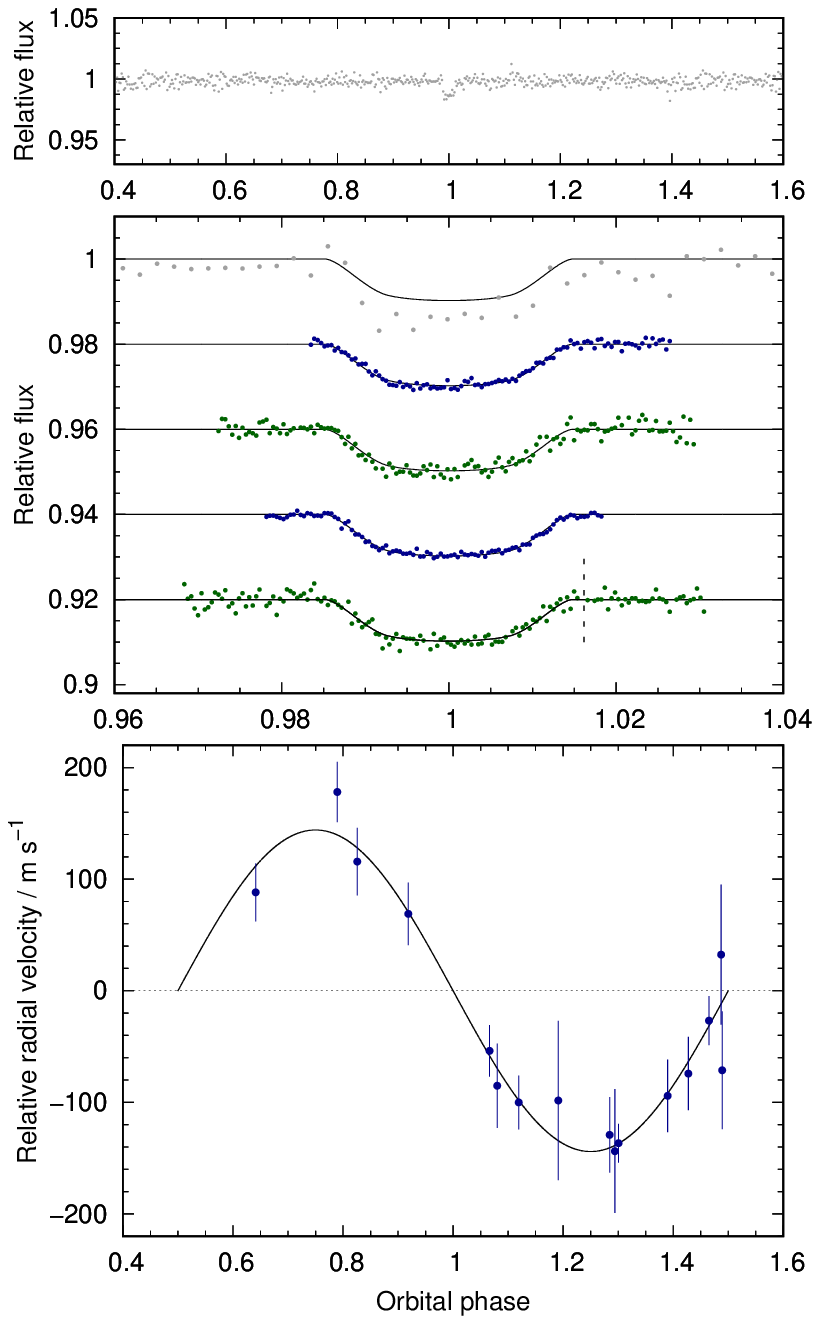}
\caption{WASP-146b discovery data. 
As for \cref{fig:w137-rv-phot}.
\label{fig:w146-rv-phot}}
\end{figure}

From periodic dimmings seen in their SuperWASP-North and WASP-South lightcurves (\citealt{2006PASP..118.1407P}; top panel of \cref{fig:w134-rv-phot,fig:w137-rv-phot,fig:w143-rv-phot,fig:w146-rv-phot}), we identified each star as a candidate host of a transiting planet using the techniques described in \citet{2006MNRAS.373..799C,2007MNRAS.380.1230C}.
We conducted photometric and spectroscopic follow-up observations using various facilities at the ESO La Silla observatory: the EulerCam imager and the CORALIE spectrograph, both mounted on the 1.2-m Swiss Euler telescope \citep{2012A&A...544A..72L, 2000A&A...354...99Q}, the 0.6-m TRAPPIST-South imager \citep{2011EPJWC..1106002G,2011Msngr.145....2J}, 
and the HARPS-S spectrograph on the 3.6-m ESO telescope \citep{2002Msngr.110....9P}. 
We obtained additional data using the HARPS-N spectrograph on the 3.6-m Telescopio Nazionale Galileo at the Observatorio del Roque de los Muchachos \citep{2012SPIE.8446E..1VC}.
We provide a summary of our observations in \cref{tab:obs}. 

\begin{deluxetable}{lcrlc}
\tabletypesize{\scriptsize}
\tablecaption{Summary of observations \label{tab:obs}}
\tablehead{
  \colhead{Facility} & \colhead{Date\tablenotemark{a}} & \colhead{$N_{\rm obs}$} & \colhead{Notes\tablenotemark{b}}
}
\startdata
{\bf WASP-134}\\
WASP	    	& 2008 Jun--2010 Oct     & 28\,614 & 400--700 nm\\
TRAPPIST-South		& 2014 Sep 05			& 678	& $I+z$			\\
TRAPPIST-South		& 2014 Oct 26			& 284	& $I+z$			\\
Euler/EulerCam		& 2016 Sep 25			& 338	& NGTS filter	\\
Euler/CORALIE	& 2014 Jul--2018 Oct		& 33	& orbit 		\\
ESO3.6/HARPS-S	& 2015 Jun--2015 Aug  		& 10 	& orbit 		\\
ESO3.6/HARPS-S	& 2015 Aug 16		  		& 17 	& transit 		\\
TNG/HARPS-N     & 2018 Aug 16				& 47	& transit		\\
{\bf WASP-137}\\
WASP	    	& 2008 Jul--2010 Dec     & 17\,463 & 400--700 nm\\
TRAPPIST-South		& 2014 Nov 10			& 577	& $I+z$			\\
Euler/EulerCam		& 2014 Nov 14			& 303	& Gunn $r$		\\
TRAPPIST-South		& 2014 Dec 27			& 709	& $I+z$			\\
TRAPPIST-South		& 2015 Sep 03			& 982	& $I+z$; MF		\\
Euler/CORALIE	& 2014 Sep--2017 Jan		& 32	& orbit 		\\
{\bf WASP-143}\\
WASP	    	& 2009 Jan--2012 Apr     & 32\,995 & 400--700 nm\\
TRAPPIST-South		& 2015 Jan 31			& 767	& blue-blocking	\\
Euler/EulerCam		& 2015 Feb 15			& 199	& NGTS filter	\\
Euler/EulerCam		& 2015 Mar 06			& 178	& NGTS filter	\\
TRAPPIST-South		& 2016 Feb 09			& 843	& blue-blocking; MF\\
Euler/CORALIE		& 2014 Feb--2017 May	& 22	& orbit 		\\
{\bf WASP-146}\\
WASP	    	& 2008 Jun--2011 Nov     & 54\,839 & 400--700 nm\\
Euler/EulerCam		& 2014 Nov 18			& 190	& NGTS filter	\\
TRAPPIST-South		& 2014 Nov 18			& 791	& blue-blocking	\\
Euler/EulerCam		& 2015 Jul 17			& 150	& NGTS filter	\\
TRAPPIST-South		& 2015 Jul 17			& 874	& blue-blocking; MF	\\
Euler/CORALIE		& 2014 Jul--2016 Oct	& 16	& orbit 		\\
\enddata
\tablenotetext{a}{The dates are `night beginning'.}
\tablenotetext{b}{For the photometry datasets, we state which filter was used. For the spectroscopy datasets, we indicate whether the data cover the orbit or the transit. `MF' indicates that TRAPPIST-South performed a meridian flip, which was accounted for by including an offset during lightcurve fitting.}
\end{deluxetable}

\begin{figure*}
\includegraphics[width=180mm]{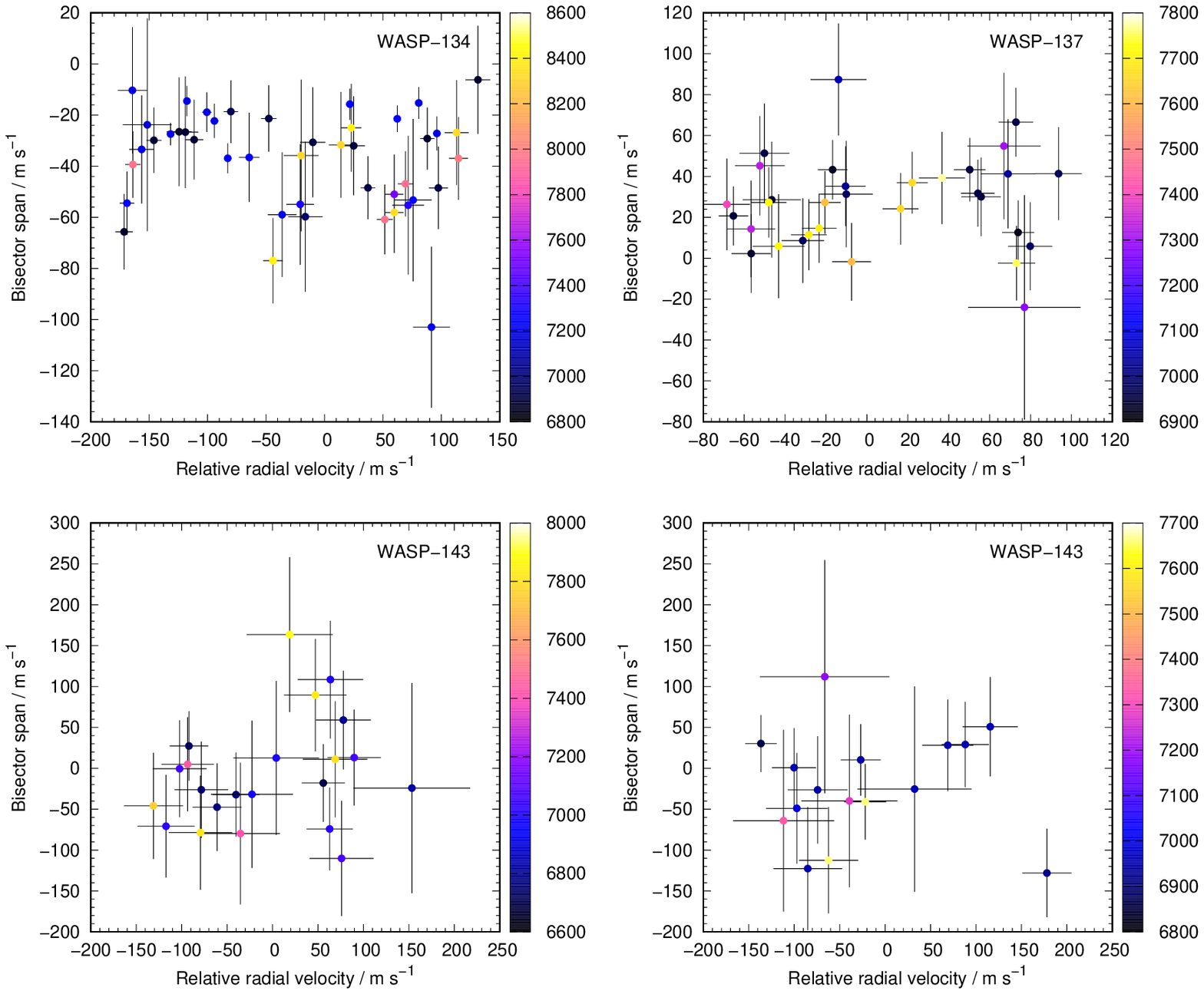}
\caption{
Bisector span versus radial velocity. The colour bar depicts the Barycentric Julian Date (BJD-2450000). 
For WASP-134, we omit the HARPS data taken through the transits as they will be affected by the RM effect.
\label{fig:bis}}
\end{figure*}

We obtained lightcurves from the time-series images using standard differential aperture photometry (second panel of \cref{fig:w134-rv-phot,fig:w137-rv-phot,fig:w143-rv-phot,fig:w146-rv-phot}). 
We computed radial-velocity (RV) measurements from the CORALIE and HARPS spectra by weighted cross-correlation with a G2 binary mask \citep{1996A&AS..119..373B,2002Msngr.110....9P}.
We detected sinusoidal variations in the RVs with semi-amplitudes consistent with planetary mass companions and that phase with the WASP ephemerides (bottom panel of \cref{fig:w134-rv-phot,fig:w137-rv-phot,fig:w143-rv-phot,fig:w146-rv-phot}). 
The lack of correlation between RV and bisector span supports our conclusion that the RV signals are induced by orbiting bodies and not by stellar activity (\cref{fig:bis}; \citealt{2001A&A...379..279Q}).

\section{Stellar analysis} \label{sec:star}
We performed spectral analyses using the procedures detailed in \cite{2013MNRAS.428.3164D} to obtain stellar effective temperature \teff, surface gravity \logg, metallicity \feh, projected rotation speed $v \sin i_{\rm *,spec}$, and lithium abundance $\log A({\rm Li})$. The results of the spectral analyses are given in \cref{tab:spec}. 
We calculated macroturbulence using the calibration of \citet{2014MNRAS.444.3592D}. 
We calculated distance using the Gaia DR2 parallax \citep{2018A&A...616A...1G}, and stellar luminosity and radius using the infrared flux method (IRFM) of \citet{1977MNRAS.180..177B}.
Using the method of \citet{2011PASP..123..547M}, we checked for modulation of the WASP lightcurves as can be caused by the combination of magnetic activity and stellar rotation. 
We find no signals with amplitudes greater than 1--2\,mmag. 

Though we can measure stellar density \densstar\ directly from the transit lightcurves, we require a constraint on stellar mass \mstar\ or radius \rstar\ for a full characterisation of the system. 
For each star we inferred \mstar\ and age $\tau$ using the {\sc bagemass} stellar evolution MCMC code of \citet{2015A&A...575A..36M}, with input of the values of \densstar\ from initial MCMC analyses (see Section~\ref{sec:mcmc}) and \teff\ and \feh\ from the spectral analyses. 
We conservatively inflated the error bar by a factor of 2 to place Gaussian priors on \mstar\ in our final MCMC analyses. 
We note that the values of \rstar\ from our final MCMC analyses are consistent with the values obtained from the IRFM and the Gaia parallax (compare the values in \cref{tab:spec,tab:mcmc}).

\begin{deluxetable*}{lcccccc}
\tablecaption{Stellar parameters \label{tab:spec}} 
\tablehead{
  \colhead{Parameter} & \colhead{Symbol} & \colhead{WASP-134} & \colhead{WASP-137} & \colhead{WASP-143} & \colhead{WASP-146} & \colhead{Unit}
}
\startdata
Constellation & \ldots & Pegasus & Cetus & Hydra & Aquarius & \ldots \\
Right Ascension (J2000) & \ldots & $\rm 21^{h} 50^{m} 16\fs77$ & $\rm 01^{h} 43^{m} 29\fs09$ & $\rm 09^{h} 23^{m} 22\fs96$ & $\rm 23^{h} 56^{m} 22\fs02$ & \ldots \\
Declination (J2000)		& \ldots & $\rm +04\degr 11\arcmin 40\fs3$ & $\rm -14\degr 08\arcmin 56\fs8$ & $\rm +02\degr 55\arcmin 57\fs1$ & $\rm -13\degr 16\arcmin 17\fs6$ & \ldots \\	
Tycho-2 $V_{\rm mag}$	& $V$ & 11.3	& 11.0 & 12.6\tablenotemark{a} & 12.9\tablenotemark{a} & \ldots \\
2MASS $K_{\rm mag}$	& $K$ & 9.4	& 9.5 & 11.3 & 11.0 & \ldots \\
Spectral type\tablenotemark{b}   & \ldots & G4 & G0 & G1 & G0 & \ldots \\
Stellar effective temperature & $T_{\rm eff}$ & 5700 $\pm$ 100 & 6100 $\pm$ 140 & 5900 $\pm$ 140 & 6100 $\pm$ 140 & K \\
Distance (Gaia) & d & $195 \pm 2$ & $289 \pm 4$ & $402 \pm 7$ & $495 \pm 27$ & pc\\
Stellar mass & \mstar & $1.131 \pm 0.045$ & $1.216 \pm 0.066$ & $1.087 \pm 0.045$ & $1.057 \pm 0.085$ & \msol \\
Stellar radius (IRFM) & $R_{\rm *,IRFM}$ & $1.16 \pm 0.06$ & $1.65 \pm 0.10$ & $1.00 \pm 0.05$ & $1.29 \pm 0.06$ & \rsol \\
Stellar surface gravity & $\log g_{*}$ & 4.4 $\pm$ 0.1 & 4.0 $\pm$ 0.2 & 4.4 $\pm$ 0.2 & 4.3 $\pm$ 0.2 & [cgs]\\
Stellar metallicity\tablenotemark{c} & [Fe/H] & $+0.40 \pm 0.07$ & $+0.08 \pm 0.07$ & $+0.23 \pm 0.10$ & $-0.01 \pm 0.16$ & \ldots\\
Stellar luminosity & $\log(L/L_\odot)$ & 0.103 $\pm$ 0.048 & 0.487 $\pm$ 0.051 & 0.027 $\pm$ 0.041 & 0.268 $\pm$ 0.044 & \ldots\\
Proj. stellar rotation speed & $v \sin i_{\rm *,spec}$ & $0.9 \pm 0.6$ & $3.7 \pm 0.9$ & $0.9 \pm 0.9$ & $0.9 \pm 0.9$ & \kms\\
Lithium abundance & $\log A({\rm Li})$ & $<0.4$ & $2.87 \pm 0.08$ & $<1.6$ & $<1.6$ & \ldots\\
Macroturbulence\tablenotemark{d} & $v_{\rm mac}$ & 3.1 & 5.0 & 3.6 & 3.8 & \kms\\
Age & $\tau$ & $5.1 \pm 1.6$ & $4.3 \pm 1.8$ & $1.9 \pm 1.5$ & $6.9 \pm 2.5$ & Gyr \\
\enddata
\tablenotetext{a}{From the USNO YB6 catalog.}
\tablenotetext{b}{Spectral type estimated using the table in \citet{1992oasp.book.....G}.}
\tablenotetext{c}{Iron abundance is relative to the solar value of \protect{\citet{2009ARAA..47..481A}}.}
\tablenotetext{d}{Macroturbulence from the calibration of \citet{2014MNRAS.444.3592D}, with an error of 0.7\,\kms.}
\end{deluxetable*} 



section{System parameters from MCMC analyses} \label{sec:mcmc}
We determined the system parameters from a simultaneous fit to the transit lightcurves and the radial velocities using the current version of the Markov-chain Monte Carlo (MCMC) code presented in \citet{2007MNRAS.380.1230C} and described further in \citet{2015A&A...575A..61A}. 
We partitioned those TRAPPIST-South lightcurves affected by meridian flips so as to account for any offsets. 
When fitting eccentric orbits, we obtained $e = 0.076 \pm 0.032$ for WASP-137b, $e = 0.0021^{+0.0014}_{-0.0007}$ for WASP-143b, and $e = 0.041^{+0.049}_{-0.029}$ for WASP-146b. 
Each value is small and of low significance, so we adopted circular orbits, as is encouraged by \citet{2012MNRAS.422.1988A} for hot Jupiters in the absence of evidence to the contrary. 

\begin{figure}
\includegraphics[width=84mm,trim={0 0.0cm 0 0},clip]{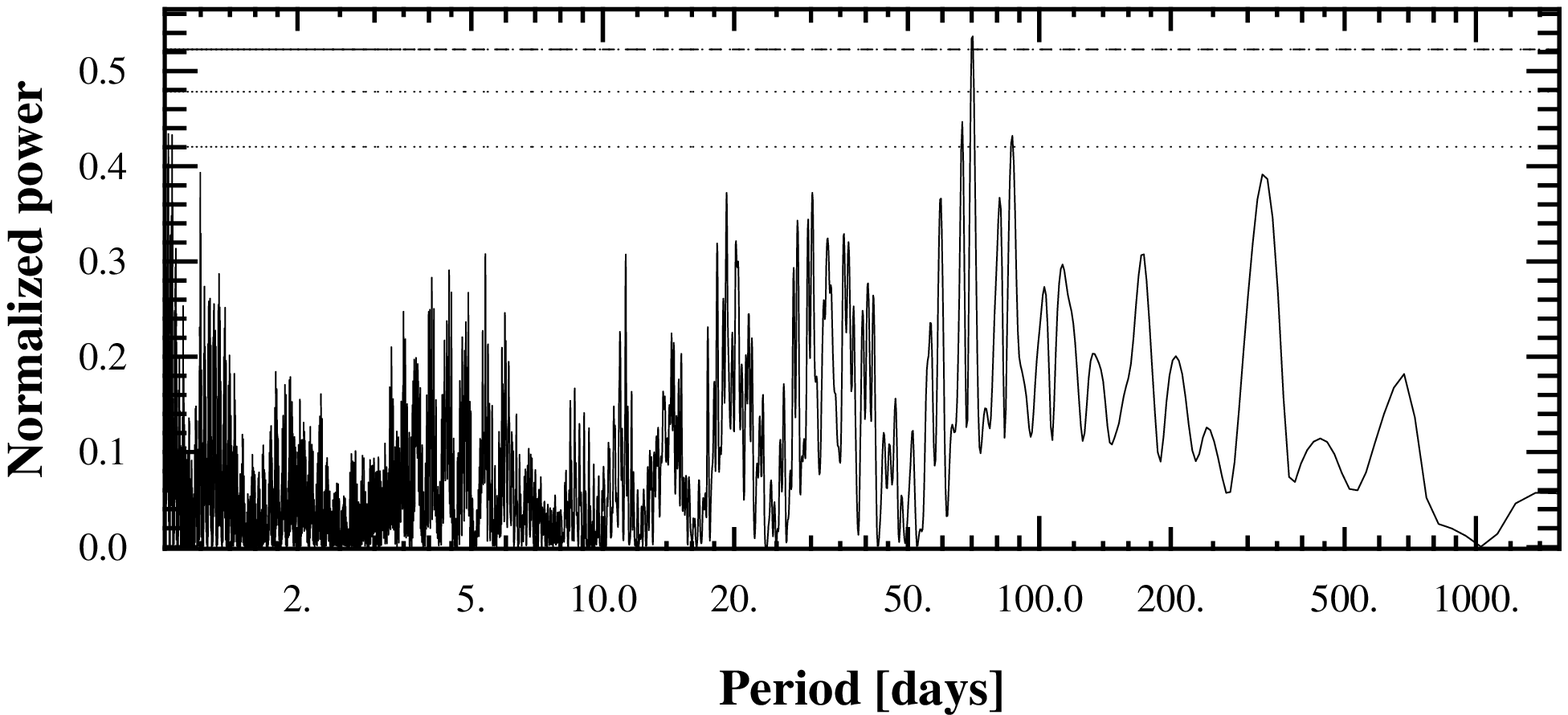}
\includegraphics[width=84mm]{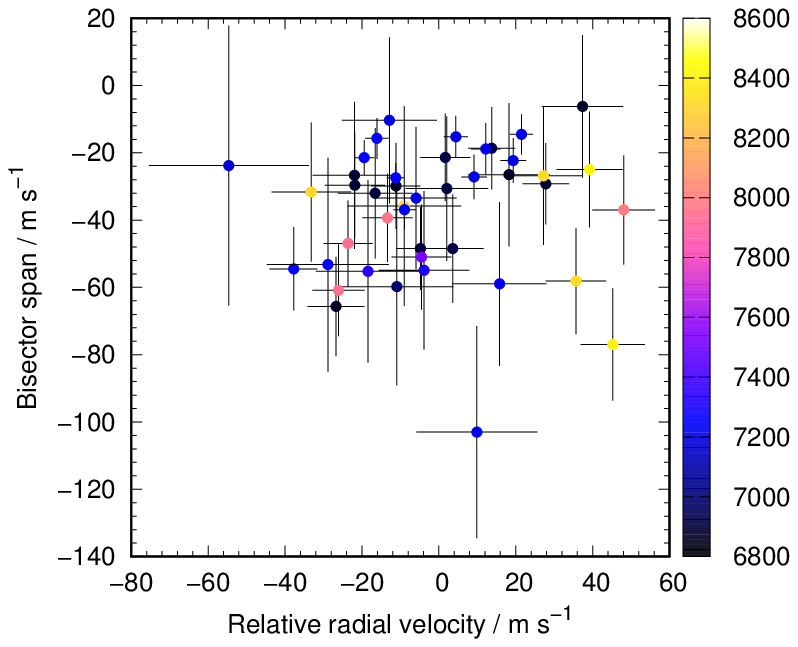}
\includegraphics[width=84mm,trim={0 0.0cm 0 1.0cm},clip]{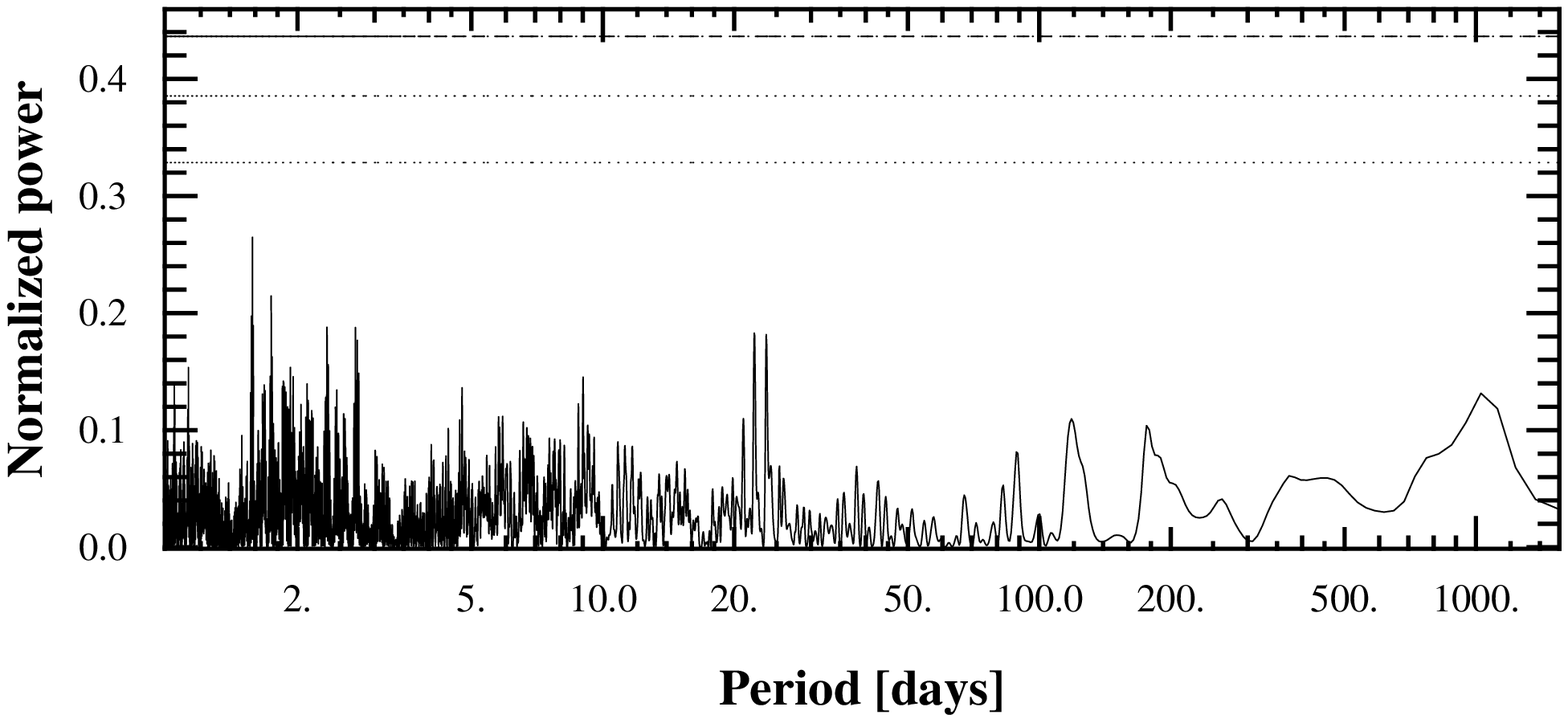}
\caption{
The evidence for WASP-134c.
{\it Top panel}: The periodogram of the residual RVs of WASP-134 after subtraction of the motion due to WASP-134b ($P = 10.15$\,d). 
The horizontal lines indicate the 10, 1 and 0.1 per cent false-alarm levels.
The peak around 70\,d, which we attribute to the planet WASP-134c, has a false-alarm probability of FAP $<$ 0.001.
{\it Middle panel}: 
Bisector span versus residual RV of WASP-134 after subtraction of the motion due to WASP-134b. The colour bar depicts the Barycentric Julian Date (BJD-2450000). 
{\it Bottom panel}: The periodogram of the residual RVs of WASP-134 after subtraction of the motion due to both WASP-134b and WASP-134c ($P = 70.0$\,d). 
\label{fig:wasp-134c}
}
\end{figure}

\begin{figure}
\includegraphics[width=84mm]{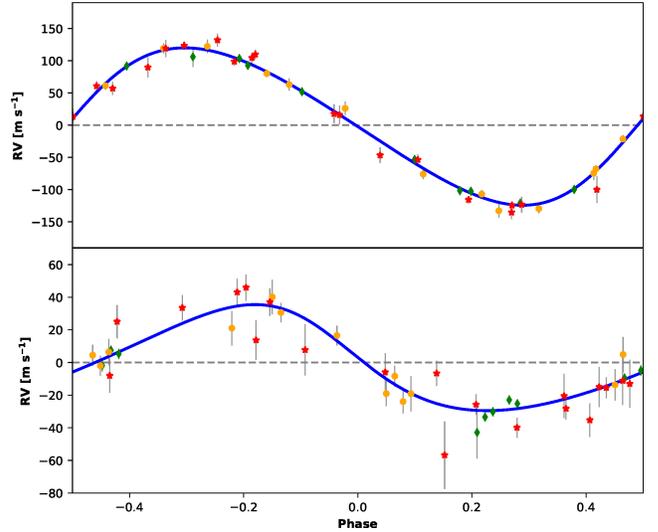}
\caption{
A two-planet fit to the RVs of WASP-134 (excluding the transit sequences). 
{\it Top panel}: The phase-folded orbit of WASP-134b ($P = 10.15$\,d).
{\it Bottom panel}: The phase-folded orbit of WASP-134c ($P = 70.0$\,d).
The CORALIE07 and CORALIE14 RVs are shown as yellow circles and red stars, respectively. The HARPS-S RVs are shown as green diamonds. 
\label{fig:rv-twoplanet}
}
\end{figure}

We adopted an eccentric orbit for WASP-134b ($P$ = 10.1\,d) as it fits the data much better than does a circular orbit ($\rm \Delta AICc = 40$). 
Having noticed excess scatter about an initial fit, we calculated a periodogram of the residuals and found a significant peak around $P$ = 70\,d (\cref{fig:wasp-134c}, top panel; FAP $<$ 0.001), which we attribute to the planet WASP-134c. 
The absence of a correlation between bisector span and residual RV (about the best-fitting orbit for WASP-134b) supports our interpretation that the 70-d signal is induced by a planet and not stellar activity (\cref{fig:wasp-134c}, middle panel).
We used the {\sc RadVel} code of \citet{2018PASP..130d4504F} to fit a two-planet model, fixing $P$ and $T_{\rm c}$ for the inner planet at the values derived from the transit photometry, placing a limit on eccentricity of $e < 0.35$, and excluding the transit sequences. 
We plot the best-fitting two-planet model in \cref{fig:rv-twoplanet} and provide the two-planet solution in \cref{tab:w134-twoplanets}. 
The two-planet model is a much better fit to the data than is the one-planet model ($\rm \Delta AICc = 111$). 
A periodogram of the residual RVs about the two-planet model shows no significant peak (\cref{fig:wasp-134c}, bottom panel). 
We do not see evidence of transits of WASP-134c in the WASP data, but the phase coverage is sparse (only a few transits of WASP-134b had good coverage). 
TESS is scheduled to observe WASP-134 during 2019 Aug 15 to 2019 Sep 11 (camera 1, Sector 15), whereas we predict the nearby inferior conjunctions of WASP-134c to occur on 2019 Aug 8 and 2019 Oct 17, with a 1-$\sigma$ uncertainty of 5\,d.
We subtracted the best-fitting orbit of WASP-134c from the RVs of WASP-134 prior to the MCMC analysis.

\begin{deluxetable}{lcccc}
\tabletypesize{\scriptsize}
\tablecaption{Two-planet solution for WASP-134\label{tab:w134-twoplanets}} 
\tablehead{
\colhead{Parameter} & \colhead{Symbol} & \colhead{WASP-134b} & \colhead{WASP-134c} & \colhead{Unit}
}
\startdata
Orbital period & $P$ & 10.1498 (fixed) & 70.01 $\pm$ 0.14 & d\\
Epoch of infer. conjunc. & $T_{\rm conj}$ & 2457464.848 (fixed)  & 2457234.2 $\pm$ 1.8 & d \\
Orbital eccentricity & $e$ & 0.146 $\pm$ 0.015 & $0.173 \pm 0.090$ & \ldots \\
Arg. of periastron & $\omega$ & $-97.2 \pm 2.6$ & $58 \pm 32$ & $^\circ$ \\
Refl. veloc. semi-ampl. & $K_{\rm 1}$ & 122.1 $\pm$ 2.1 & 32.5 $\pm$ 2.7 & \kms \\
Minimum planet mass & \mplanet\ $\sin i$ & $1.40 \pm 0.08$ & $0.70 \pm 0.07$ & \mjup \\
\enddata
\end{deluxetable}

For each system, to for instrumental and astrophysical offsets, we partitioned the RV datasets and fit a separate systemic velocity to each of them. This included the CORALIE RVs from before and after the November 2014 upgrade (labelled CORALIE07 and CORALIE14, respectively), and the HARPS RVs around the orbit and through the transits of WASP-134b.

For WASP-134b, we modelled the Rossiter-McLaughlin (RM) effect using the formulation of \citet{2011ApJ...742...69H}. Due to the low transit impact parameter, there is a degeneracy between the projected stellar rotation speed $v \sin i_{\rm *,RM}$ and the projected stellar obliquity $\lambda$ (e.g. \citealt{2011ApJ...738...50A}). This results in values of $v \sin i_{\rm *,RM}$ ($>$ 9\,\kms) far higher than the value from our spectral analysis ($v \sin i_{\rm *,spec} = 0.9 \pm 0.6$\,\kms), so we placed a Gaussian prior on $v \sin i_{\rm *,RM}$ using the value from the spectral analysis. 
With the prior, we obtained $\lambda = -43.7 \pm 9.9^\circ$ and $v \sin i_{\rm *,RM} = 2.08 \pm 0.26$\,\kms. 
Without the prior, we obtained $\lambda = -67^{+15}_{-9}$$^\circ$ and $v \sin i_{\rm *,RM} = 3.3^{+1.5}_{-0.9}$\,\kms. 

\begin{figure*}
\includegraphics[width=180mm]{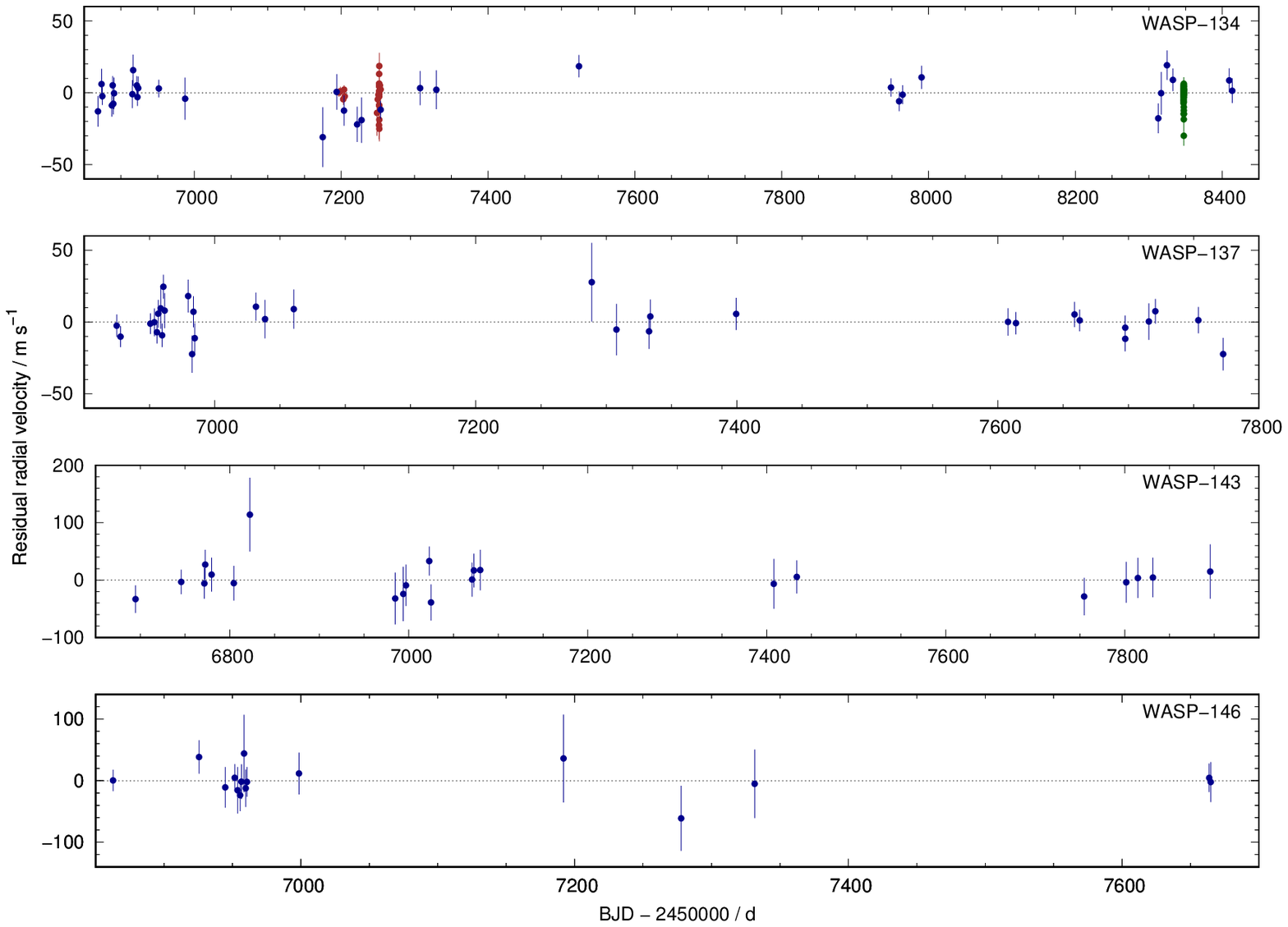}
\caption{
The residual RVs about the best-fitting orbital and RM effect models. The symbol colours are the same as in \cref{fig:w134-rv-phot}.
\label{fig:rv-time}}
\end{figure*}

We present the median values and 1-$\sigma$ limits on the system parameters from our final MCMC analyses in \cref{tab:mcmc}. We plot the best fits to the RVs and the transit lightcurves in \cref{fig:w134-rv-phot,fig:w137-rv-phot,fig:w143-rv-phot,fig:w146-rv-phot} and the residuals of the RVs about the best-fitting orbital models in \cref{fig:rv-time}.

\begin{deluxetable*}{lcccccc}
\tabletypesize{\scriptsize}
\tablecaption{System parameters\label{tab:mcmc}} 
\tablehead{
\colhead{Parameter} & \colhead{Symbol} & \colhead{WASP-134\tablenotemark{a}} & \colhead{WASP-137} & \colhead{WASP-143} & \colhead{WASP-146} & \colhead{Unit}
}
\startdata
\\
\multicolumn{7}{l}{\it MCMC Gaussian priors}\\
Stellar mass & $M_{\rm *}$ & 1.13 $\pm$ 0.09 & 1.22 $\pm$ 0.13 & 1.087 $\pm$ 0.090 & 1.06 $\pm$ 0.17 & $M_{\rm \odot}$ \\
Stellar effective temperature & $T_{\rm eff}$ & 5700 $\pm$ 100 & 6100 $\pm$ 140 & 5900 $\pm$ 1400 & 5900 $\pm$ 140 & K \\
\\
\multicolumn{7}{l}{\it MCMC parameters controlled by Gaussian priors}\\
Stellar mass & $M_{\rm *}$ & 1.130 $\pm$ 0.091 & 1.22 $\pm$ 0.13 & 1.096 $\pm$ 0.091 & 1.06 $\pm$ 0.17 & $M_{\rm \odot}$ \\
Stellar effective temperature & $T_{\rm eff}$ & 5574 $\pm$ 99 & 6127 $\pm$ 136 & 6042 $\pm$ 135 & 5894 $\pm$ 140 & K \\
\\
\multicolumn{7}{l}{\it MCMC fitted parameters}\\
Orbital period & $P$ & 10.1467583 $\pm$ 0.0000080 & 3.9080284 $\pm$ 0.0000053 & 3.7788730 $\pm$ 0.0000032 & 3.3969440 $\pm$ 0.0000036 & d\\
Transit epoch (HJD) & $T_{\rm c}$ & 2457201.03099 $\pm$ 0.00075 & 2456937.61342 $\pm$ 0.00106 & 2457099.94471 $\pm$ 0.00015 & 2457109.72182 $\pm$ 0.00021 & d\\
Transit duration & $T_{\rm 14}$ & 0.2218 $\pm$ 0.0019 & 0.1425 $\pm$ 0.0034 & 0.12858 $\pm$ 0.00055 & 0.09980 $\pm$ 0.00097 & d\\
Planet-to-star area ratio & $R_{\rm P}^{2}$/R$_{*}^{2}$ & 0.00749 $\pm$ 0.00020 & 0.00737 $\pm$ 0.00031 & 0.01569 $\pm$ 0.00017 & 0.01049 $\pm$ 0.00017 & \ldots \\
Impact parameter\tablenotemark{b} & $b$ & 0.306 $\pm$ 0.082 & 0.690 $\pm$ 0.049 & 0.181 $\pm$ 0.097 & 0.8290 $\pm$ 0.0089 & \ldots \\
Reflex velocity semi-amplitude & $K_{\rm 1}$ & 0.1220 $\pm$ 0.0012 & 0.0767 $\pm$ 0.0026 & 0.0890 $\pm$ 0.0085 & 0.144 $\pm$ 0.012 & \kms\\
Systemic velocity (CORALIE07) & $\gamma_{\rm CORALIE07}$ & 5.7967 $\pm$ 0.0021 & 4.6948 $\pm$ 0.0027 & 20.215 $\pm$ 0.010 & $-$5.6464 $\pm$ 0.0084 & \kms\\
Systemic velocity (CORALIE14) & $\gamma_{\rm CORALIE14}$ & 5.7975 $\pm$ 0.0020 & 4.6937 $\pm$ 0.0022 & 20.2008 $\pm$ 0.0088 & $-$5.6144 $\pm$ 0.015 & \kms\\
Systemic velocity (HARPS-S) & $\gamma_{\rm HARPS-S}$ & 5.82223 $\pm$ 0.00097 & \ldots & \ldots & \ldots & \kms\\
Systemic velocity (HARPS-S RM) & $\gamma_{\rm HARPS-S_{\rm RM}}$ & 5.8200 $\pm$ 0.0015 & \ldots & \ldots & \ldots & \kms\\
Systemic velocity (HARPS-N RM) & $\gamma_{\rm HARPS-N_{\rm RM}}$ & 5.83155 $\pm$ 0.00048 & \ldots & \ldots & \ldots & \kms\\
First eccentricity parameter\tablenotemark{c} & \ecos & $-0.0187 \pm 0.0042$ & \ldots & \ldots & \ldots & \\
Second eccentricity parameter\tablenotemark{c} & \esin & $-0.1435 \pm 0.0086$ & \ldots & \ldots & \ldots & \\
First obliquity parameter\tablenotemark{c} & \vsicos & $1.49 \pm 0.20$ & \ldots & \ldots & \ldots & \\
Second obliquity parameter\tablenotemark{c} & \vsisin & $-1.42 \pm 0.39$ & \ldots & \ldots & \ldots & \\
\\
\multicolumn{7}{l}{\it MCMC derived parameters}\\
Orbital eccentricity & $e$ & 0.1447 $\pm$ 0.0086 ($<$ 0.16 at 2$\sigma$) & 0\tablenotemark{d} ($<$ 0.14 at 2$\sigma$) & 0\tablenotemark{d} ($<$ 0.0007 at 2$\sigma$) & 0\tablenotemark{d} ($<$ 0.15 at 2$\sigma$) & \ldots \\
Argument of periastron & $\omega$ & $-97.4 \pm 1.7$ & \ldots & \ldots & \ldots & $^\circ$ \\
Sky-projected stellar obliquity & $\lambda$ & $-$43.7 $\pm$ 9.9 & \ldots & \ldots & \ldots & $^\circ$ \\
Sky-projected stellar rotation speed & $v \sin i_{\rm *,RM}$ & 2.08 $\pm$ 0.26 & \ldots & \ldots & \ldots & \kms \\
Scaled semi-major axis & $a/R_{\rm *}$ & $17.53 \pm 0.53$ & $7.31 \pm 0.47$ & $10.39 \pm 0.17$ & $7.88 \pm 0.16$ & \ldots \\
Orbital inclination & $i$ & 89.13 $\pm$ 0.26 & 84.59 $\pm$ 0.73 & 89.00 $\pm$ 0.55 & 83.96 $\pm$ 0.19 & $^\circ$\\
Ingress and egress duration & $T_{\rm 12}=T_{\rm 34}$ & 0.0193 $\pm$ 0.0013 & 0.0203 $\pm$ 0.0030 & 0.01473 $\pm$ 0.00056 & 0.0263 $\pm$ 0.0015 & d\\
Stellar radius & $R_{\rm *}$ & $1.175 \pm 0.048$ & 1.52 $\pm$ 0.11 & 1.013 $\pm$ 0.032 & 1.232 $\pm$ 0.072 & $R_{\rm \odot}$\\
Stellar surface gravity & $\log g_{*}$ & 4.352 $\pm$ 0.029 & 4.155 $\pm$ 0.059 & 4.465 $\pm$ 0.018 & 4.282 $\pm$ 0.029 & [cgs]\\
Stellar density & $\rho_{\rm *}$ & 0.702 $\pm$ 0.063 & 0.343 $\pm$ 0.066 & 1.054 $\pm$ 0.050 & 0.568 $\pm$ 0.035 & $\rho_{\rm \odot}$\\
Planetary mass & $M_{\rm P}$ & $1.412 \pm 0.075$ & 0.681 $\pm$ 0.054 & 0.725 $\pm$ 0.084 & 1.11 $\pm$ 0.15 & $M_{\rm Jup}$\\
Planetary radius & $R_{\rm P}$ & 0.988 $\pm$ 0.057 &  1.27 $\pm$ 0.11 & 1.234 $\pm$ 0.042 & 1.228 $\pm$ 0.076 & $R_{\rm Jup}$\\
Planetary surface gravity & $\log g_{\rm P}$ & 3.521 $\pm$ 0.034 & 2.983 $\pm$ 0.069 & 3.033 $\pm$ 0.045 & 3.229 $\pm$ 0.044 & [cgs]\\
Planetary density & $\rho_{\rm P}$ & 1.47 $\pm$ 0.18 & 0.333 $\pm$ 0.081 & 0.382 $\pm$ 0.045 & 0.604 $\pm$ 0.084 & $\rho_{\rm J}$\\
Orbital semi-major axis & $a$ & 0.0956 $\pm$ 0.0025 & 0.0519 $\pm$ 0.0018 & 0.0490 $\pm$ 0.0014 & 0.0451 $\pm$ 0.0024 & AU\\
Planetary equilibrium temperature\tablenotemark{e} & $T_{\rm eql}$ & 953 $\pm$ 22 & 1601 $\pm$ 65 & 1325 $\pm$ 30 & 1486 $\pm$ 43 & K\\
\enddata
\tablenotetext{a}{These values are for WASP-134b. For the values relating to WASP-134c, see \cref{tab:w134-twoplanets}.}
\tablenotetext{b}{Impact parameter is the distance between the centre of the stellar disc and the transit chord: $b = a \cos i / R_{\rm *}$.}
\tablenotetext{c}{We actually fit \secos, \sesin, \svsicos\ and \svsisin, but we give these quantities for ease of interpretation and comparison with other studies.}
\tablenotetext{d}{We assumed circular orbits for these systems.}
\tablenotetext{e}{Equilibrium temperature calculated assuming zero albedo and efficient redistribution of heat from the planet's presumed permanent day-side to its night-side.}
\end{deluxetable*} 

\newpage

\section{Discussion} \label{sec:disc}
We have presented the discovery of five Jupiter-mass planets (\mplanet\ = 0.68--1.41\,\mjup) orbiting moderately bright ($V$ = 11.0--12.9) Solar-type stars (\mstar\ = 1.06--1.22\,\msol). 
As fairly typical hot Jupiters (\teql\ = 1300--1600\,K) with orbital periods of $P$ = 3--4\,d, WASP-137b, WASP-143b and WASP-146b are remarkably similar to each other (\cref{tab:spec,tab:mcmc}). 

The WASP-134 system is rather more interesting.
WASP-134 is a metal-rich G4 star ([Fe/H] = +0.40 $\pm$ 0.07) orbited by two warm Jupiters. WASP-134b (\mplanet\ = 1.41\,\mjup; \teql\ = 950\,K) is in an eccentric ($e = 0.15 \pm 0.01$), 10.15-d orbit ($a$ = 0.096\,AU) that is misaligned with the spin of the star ($\lambda = -44 \pm 10^\circ$). Its companion, WASP-134c (\mplanet\ $\sin i$ = 0.70\,\mjup; \teql\ = 500\,K), is in an eccentric ($e = 0.17 \pm 0.09$), 70.0-d orbit ($a$ = 0.35\,AU).
Thus WASP-134 is a rare type of system: a hot/warm Jupiter with a nearby giant companion. 
In that respect, WASP-134 may be similar to the HAT-P-46 system. \citet{2014AJ....147..128H} found that their RVs of HAT-P-46 are fit well with a two-planet model: \mplanet\ = 0.49 and $P = 4.5$\,d for HAT-P-46b and \mplanet\ $\sin i$ = 2.0\,\mjup\ and $P = 78$\,d for the candidate planet HAT-P-46c. 
The evidence for HAT-P-46c, though, is not yet conclusive and further RV monitoring is required. 

Of those hot/warm Jupiters with confirmed giant companions, the companions tend to be very far out (e.g. HAT-P-17c with $P = 1610$\,d; \citealt{2012ApJ...749..134H}), and rarely closer than 1\,AU (e.g. WASP-41c at $a = 1.07$\,AU and WASP-47c at $a = 1.36$\,AU; \citealt{2016A&A...586A..93N}). 
Also of note are those hot-Jupiter systems with brown-dwarf companions within a few AU, such as WASP-53 and WASP-81 \citep{2017MNRAS.467.1714T}.
In their tabulation of hot and warm Jupiters with nearby giant companions, \citet{2016AJ....152..174A} list only three companions inside of 1\,AU, the closest of which is  HD\,9446\,c at 0.65\,AU \citep{2010A&A...513A..69H}. With $a = 0.35$\,AU, WASP-134c is in a much shorter orbit. 

It seems unlikely that WASP-134b could have arrived in situ via high-eccentricity migration (e.g \citealt{2016ApJ...829..132P}). 
\citet{2016AJ....152..174A} studied the observed population of hot and warm Jupiters with nearby giant companions and found that the ejection of a planet or its collision with the star are more likely outcomes when exploring such pathways. Thus in-situ formation (e.g. \citealt{2016ApJ...825...98H}) or disc migration (e.g. \citealt{1996Natur.380..606L}) seem more likely explanations.

\acknowledgments
SuperWASP-North is hosted by the Issac Newton Group on La Palma and WASP-South is hosted by SAAO; we are grateful for their support and assistance.
Funding for WASP comes from consortium universities and from the UK's Science and Technology Facilities Council. 
The Swiss {\it Euler} Telescope is operated by the University of Geneva, and is funded by the Swiss National Science Foundation. 
The research leading to these results has received funding from the  ARC grant for Concerted Research Actions, financed by the Wallonia-Brussels Federation. TRAPPIST is funded by the Belgian Fund for Scientific Research (Fond National de la Recherche Scientifique, FNRS) under the grant FRFC 2.5.594.09.F, with the participation of the Swiss National Science Foundation (SNF). MG and EJ are FNRS Senior Research Associates.
Based on observations collected at the European Organisation for Astronomical Research in the Southern Hemisphere under ESO programmes 095.C-0105(A) and 097.C-0434(B).
Based on observations under programme CAT18A\_138 made with the Italian Telescopio Nazionale Galileo (TNG) operated on the island of La Palma by the Fundaci\'on Galileo Galilei of the INAF (Istituto Nazionale di Astrofisica) at the Spanish Observatorio del Roque de los Muchachos of the Instituto de Astrofisica de Canarias. 
This research has made use of TEPCat, a catalogue of the physical properties of transiting planetary systems maintained by John Southworth.

\vspace{5mm}
\facilities{SuperWASP (SuperWASP-North, WASP-South), TRAPPIST, Euler1.2m (CORALIE, EulerCam), ESO:3.6m (HARPS-S), TNG (HARPS-N)}


\begin{thebibliography}{}
\expandafter\ifx\csname natexlab\endcsname\relax\def\natexlab#1{#1}\fi
\providecommand{\url}[1]{\href{#1}{#1}}
\providecommand{\dodoi}[1]{doi:~\href{http://doi.org/#1}{\nolinkurl{#1}}}
\providecommand{\doeprint}[1]{\href{http://ascl.net/#1}{\nolinkurl{http://ascl.net/#1}}}
\providecommand{\doarXiv}[1]{\href{https://arxiv.org/abs/#1}{\nolinkurl{https://arxiv.org/abs/#1}}}

\bibitem[{{Albrecht} {et~al.}(2011){Albrecht}, {Winn}, {Johnson}, {Butler},
  {Crane}, {Shectman}, {Thompson}, {Narita}, {Sato}, {Hirano}, {Enya}, \&
  {Fischer}}]{2011ApJ...738...50A}
{Albrecht}, S., {Winn}, J.~N., {Johnson}, J.~A., {et~al.} 2011, \apj, 738, 50,
  \dodoi{10.1088/0004-637X/738/1/50}

\bibitem[{{Anderson} {et~al.}(2012){Anderson}, {Collier Cameron}, {Gillon},
  {Hellier}, {Jehin}, {Lendl}, {Maxted}, {Queloz}, {Smalley}, {Smith},
  {Triaud}, {West}, {Pepe}, {Pollacco}, {S{\'e}gransan}, {Todd}, \&
  {Udry}}]{2012MNRAS.422.1988A}
{Anderson}, D.~R., {Collier Cameron}, A., {Gillon}, M., {et~al.} 2012, \mnras,
  422, 1988, \dodoi{10.1111/j.1365-2966.2012.20635.x}

\bibitem[{{Anderson} {et~al.}(2015{\natexlab{a}}){Anderson}, {Triaud},
  {Turner}, {Brown}, {Clark}, {Smalley}, {Collier Cameron}, {Doyle}, {Gillon},
  {Hellier}, {Lovis}, {Maxted}, {Pollacco}, {Queloz}, \&
  {Smith}}]{2015ApJ...800L...9A}
{Anderson}, D.~R., {Triaud}, A.~H.~M.~J., {Turner}, O.~D., {et~al.}
  2015{\natexlab{a}}, \apjl, 800, L9, \dodoi{10.1088/2041-8205/800/1/L9}

\bibitem[{{Anderson} {et~al.}(2015{\natexlab{b}}){Anderson}, {Collier Cameron},
  {Hellier}, {Lendl}, {Lister}, {Maxted}, {Queloz}, {Smalley}, {Smith},
  {Triaud}, {Brown}, {Gillon}, {Neveu-VanMalle}, {Pepe}, {Pollacco},
  {S{\'e}gransan}, {Udry}, {West}, \& {Wheatley}}]{2015A&A...575A..61A}
{Anderson}, D.~R., {Collier Cameron}, A., {Hellier}, C., {et~al.}
  2015{\natexlab{b}}, \aap, 575, A61, \dodoi{10.1051/0004-6361/201423591}

\bibitem[{{Antonini} {et~al.}(2016){Antonini}, {Hamers}, \&
  {Lithwick}}]{2016AJ....152..174A}
{Antonini}, F., {Hamers}, A.~S., \& {Lithwick}, Y. 2016, \aj, 152, 174,
  \dodoi{10.3847/0004-6256/152/6/174}

\bibitem[{{Asplund} {et~al.}(2009){Asplund}, {Grevesse}, {Sauval}, \&
  {Scott}}]{2009ARAA..47..481A}
{Asplund}, M., {Grevesse}, N., {Sauval}, A.~J., \& {Scott}, P. 2009, \araa, 47,
  481, \dodoi{10.1146/annurev.astro.46.060407.145222}

\bibitem[{{Bakos}(2018)}]{2018haex.bookE.111B}
{Bakos}, G.~{\'A}. 2018, {The HATNet and HATSouth Exoplanet Surveys}, 111

\bibitem[{{Baranne} {et~al.}(1996){Baranne}, {Queloz}, {Mayor}, {Adrianzyk},
  {Knispel}, {Kohler}, {Lacroix}, {Meunier}, {Rimbaud}, \&
  {Vin}}]{1996A&AS..119..373B}
{Baranne}, A., {Queloz}, D., {Mayor}, M., {et~al.} 1996, \aaps, 119, 373

\bibitem[{{Becker} {et~al.}(2015){Becker}, {Vanderburg}, {Adams}, {Rappaport},
  \& {Schwengeler}}]{2015ApJ...812L..18B}
{Becker}, J.~C., {Vanderburg}, A., {Adams}, F.~C., {Rappaport}, S.~A., \&
  {Schwengeler}, H.~M. 2015, \apjl, 812, L18,
  \dodoi{10.1088/2041-8205/812/2/L18}

\bibitem[{{Blackwell} \& {Shallis}(1977)}]{1977MNRAS.180..177B}
{Blackwell}, D.~E., \& {Shallis}, M.~J. 1977, \mnras, 180, 177

\bibitem[{{Broeg} {et~al.}(2013){Broeg}, {Fortier}, {Ehrenreich}, {Alibert},
  {Baumjohann}, {Benz}, {Deleuil}, {Gillon}, {Ivanov}, {Liseau}, {Meyer},
  {Oloffson}, {Pagano}, {Piotto}, {Pollacco}, {Queloz}, {Ragazzoni}, {Renotte},
  {Steller}, \& {Thomas}}]{2013EPJWC..4703005B}
{Broeg}, C., {Fortier}, A., {Ehrenreich}, D., {et~al.} 2013, in European
  Physical Journal Web of Conferences, Vol.~47, European Physical Journal Web
  of Conferences, 03005

\bibitem[{{Collier Cameron} {et~al.}(2006){Collier Cameron}, {Pollacco},
  {Street}, {Lister}, {West}, {Wilson}, {Pont}, {Christian}, {Clarkson},
  {Enoch}, {Evans}, {Fitzsimmons}, {Haswell}, {Hellier}, {Hodgkin}, {Horne},
  {Irwin}, {Kane}, {Keenan}, {Norton}, {Parley}, {Osborne}, {Ryans}, {Skillen},
  \& {Wheatley}}]{2006MNRAS.373..799C}
{Collier Cameron}, A., {Pollacco}, D., {Street}, R.~A., {et~al.} 2006, \mnras,
  373, 799, \dodoi{10.1111/j.1365-2966.2006.11074.x}

\bibitem[{{Collier Cameron} {et~al.}(2007){Collier Cameron}, {Wilson}, {West},
  {Hebb}, {Wang}, {Aigrain}, {Bouchy}, {Christian}, {Clarkson}, {Enoch},
  {Esposito}, {Guenther}, {Haswell}, {H{\'e}brard}, {Hellier}, {Horne},
  {Irwin}, {Kane}, {Loeillet}, {Lister}, {Maxted}, {Mayor}, {Moutou}, {Parley},
  {Pollacco}, {Pont}, {Queloz}, {Ryans}, {Skillen}, {Street}, {Udry}, \&
  {Wheatley}}]{2007MNRAS.380.1230C}
{Collier Cameron}, A., {Wilson}, D.~M., {West}, R.~G., {et~al.} 2007, \mnras,
  380, 1230, \dodoi{10.1111/j.1365-2966.2007.12195.x}

\bibitem[{{Cooke} {et~al.}(2018){Cooke}, {Pollacco}, {West}, {McCormac}, \&
  {Wheatley}}]{2018A&A...619A.175C}
{Cooke}, B.~F., {Pollacco}, D., {West}, R., {McCormac}, J., \& {Wheatley},
  P.~J. 2018, \aap, 619, A175, \dodoi{10.1051/0004-6361/201834014}

\bibitem[{{Cosentino} {et~al.}(2012){Cosentino}, {Lovis}, {Pepe}, {Collier
  Cameron}, {Latham}, {Molinari}, {Udry}, {Bezawada}, {Black}, {Born},
  {Buchschacher}, {Charbonneau}, {Figueira}, {Fleury}, {Galli}, {Gallie},
  {Gao}, {Ghedina}, {Gonzalez}, {Gonzalez}, {Guerra}, {Henry}, {Horne},
  {Hughes}, {Kelly}, {Lodi}, {Lunney}, {Maire}, {Mayor}, {Micela}, {Ordway},
  {Peacock}, {Phillips}, {Piotto}, {Pollacco}, {Queloz}, {Rice}, {Riverol},
  {Riverol}, {San Juan}, {Sasselov}, {Segransan}, {Sozzetti}, {Sosnowska},
  {Stobie}, {Szentgyorgyi}, {Vick}, \& {Weber}}]{2012SPIE.8446E..1VC}
{Cosentino}, R., {Lovis}, C., {Pepe}, F., {et~al.} 2012, in Society of
  Photo-Optical Instrumentation Engineers (SPIE) Conference Series, Vol. 8446,
  Society of Photo-Optical Instrumentation Engineers (SPIE) Conference Series

\bibitem[{{Delrez} {et~al.}(2018){Delrez}, {Gillon}, {Queloz}, {Demory},
  {Almleaky}, {de Wit}, {Jehin}, {Triaud}, {Barkaoui}, {Burdanov}, {Burgasser},
  {Ducrot}, {McCormac}, {Murray}, {Silva Fernandes}, {Sohy}, {Thompson}, {Van
  Grootel}, {Alonso}, {Benkhaldoun}, \& {Rebolo}}]{2018SPIE10700E..1ID}
{Delrez}, L., {Gillon}, M., {Queloz}, D., {et~al.} 2018, in Society of
  Photo-Optical Instrumentation Engineers (SPIE) Conference Series, Vol. 10700,
  Ground-based and Airborne Telescopes VII, 107001I

\bibitem[{{Doyle} {et~al.}(2014){Doyle}, {Davies}, {Smalley}, {Chaplin}, \&
  {Elsworth}}]{2014MNRAS.444.3592D}
{Doyle}, A.~P., {Davies}, G.~R., {Smalley}, B., {Chaplin}, W.~J., \&
  {Elsworth}, Y. 2014, \mnras, 444, 3592, \dodoi{10.1093/mnras/stu1692}

\bibitem[{{Doyle} {et~al.}(2013){Doyle}, {Smalley}, {Maxted}, {Anderson},
  {Cameron}, {Gillon}, {Hellier}, {Pollacco}, {Queloz}, {Triaud}, \&
  {West}}]{2013MNRAS.428.3164D}
{Doyle}, A.~P., {Smalley}, B., {Maxted}, P.~F.~L., {et~al.} 2013, \mnras, 428,
  3164, \dodoi{10.1093/mnras/sts267}

\bibitem[{{Fulton} {et~al.}(2018){Fulton}, {Petigura}, {Blunt}, \&
  {Sinukoff}}]{2018PASP..130d4504F}
{Fulton}, B.~J., {Petigura}, E.~A., {Blunt}, S., \& {Sinukoff}, E. 2018, \pasp,
  130, 044504, \dodoi{10.1088/1538-3873/aaaaa8}

\bibitem[{{Gaia Collaboration} {et~al.}(2018){Gaia Collaboration}, {Brown},
  {Vallenari}, {Prusti}, {de Bruijne}, {Babusiaux}, {Bailer-Jones}, {Biermann},
  {Evans}, {Eyer}, \& et~al.}]{2018A&A...616A...1G}
{Gaia Collaboration}, {Brown}, A.~G.~A., {Vallenari}, A., {et~al.} 2018, \aap,
  616, A1, \dodoi{10.1051/0004-6361/201833051}

\bibitem[{{Gillon} {et~al.}(2011){Gillon}, {Jehin}, {Magain}, {Chantry},
  {Hutsem{\'e}kers}, {Manfroid}, {Queloz}, \& {Udry}}]{2011EPJWC..1106002G}
{Gillon}, M., {Jehin}, E., {Magain}, P., {et~al.} 2011, Detection and Dynamics
  of Transiting Exoplanets, St.~Michel l'Observatoire, France, Edited by
  F.~Bouchy; R.~D{\'{\i}}az; C.~Moutou; EPJ Web of Conferences, Volume 11,
  id.06002, 11, 6002, \dodoi{10.1051/epjconf/20101106002}

\bibitem[{{Gray}(1992)}]{1992oasp.book.....G}
{Gray}, D.~F. 1992, {The observation and analysis of stellar photospheres.}

\bibitem[{{Hartman} {et~al.}(2014){Hartman}, {Bakos}, {Torres}, {Kov{\'a}cs},
  {Johnson}, {Howard}, {Marcy}, {Latham}, {Bieryla}, {Buchhave}, {Bhatti},
  {B{\'e}ky}, {Csubry}, {Penev}, {de Val-Borro}, {Noyes}, {Fischer},
  {Esquerdo}, {Everett}, {Szklen{\'a}r}, {Zhou}, {Bayliss}, {Shporer},
  {Fulton}, {Sanchis-Ojeda}, {Falco}, {L{\'a}z{\'a}r}, {Papp}, \&
  {S{\'a}ri}}]{2014AJ....147..128H}
{Hartman}, J.~D., {Bakos}, G.~{\'A}., {Torres}, G., {et~al.} 2014, \aj, 147,
  128, \dodoi{10.1088/0004-6256/147/6/128}

\bibitem[{{H{\'e}brard} {et~al.}(2010){H{\'e}brard}, {Bonfils},
  {S{\'e}gransan}, {Moutou}, {Delfosse}, {Bouchy}, {Boisse}, {Arnold},
  {Desort}, {D{\'{\i}}az}, {Eggenberger}, {Ehrenreich}, {Forveille},
  {Lagrange}, {Lovis}, {Pepe}, {Perrier}, {Pont}, {Queloz}, {Santos}, {Udry},
  \& {Vidal-Madjar}}]{2010A&A...513A..69H}
{H{\'e}brard}, G., {Bonfils}, X., {S{\'e}gransan}, D., {et~al.} 2010, \aap,
  513, A69, \dodoi{10.1051/0004-6361/200913790}

\bibitem[{{Hellier} {et~al.}(2012){Hellier}, {Anderson}, {Collier Cameron},
  {Doyle}, {Fumel}, {Gillon}, {Jehin}, {Lendl}, {Maxted}, {Pepe}, {Pollacco},
  {Queloz}, {S{\'e}gransan}, {Smalley}, {Smith}, {Southworth}, {Triaud},
  {Udry}, \& {West}}]{2012MNRAS.426..739H}
{Hellier}, C., {Anderson}, D.~R., {Collier Cameron}, A., {et~al.} 2012, \mnras,
  426, 739, \dodoi{10.1111/j.1365-2966.2012.21780.x}

\bibitem[{{Hirano} {et~al.}(2011){Hirano}, {Suto}, {Winn}, {Taruya}, {Narita},
  {Albrecht}, \& {Sato}}]{2011ApJ...742...69H}
{Hirano}, T., {Suto}, Y., {Winn}, J.~N., {et~al.} 2011, \apj, 742, 69,
  \dodoi{10.1088/0004-637X/742/2/69}

\bibitem[{{Howard} {et~al.}(2012){Howard}, {Bakos}, {Hartman}, {Torres},
  {Shporer}, {Mazeh}, {Kov{\'a}cs}, {Latham}, {Noyes}, {Fischer}, {Johnson},
  {Marcy}, {Esquerdo}, {B{\'e}ky}, {Butler}, {Sasselov}, {Stefanik},
  {Perumpilly}, {L{\'a}z{\'a}r}, {Papp}, \& {S{\'a}ri}}]{2012ApJ...749..134H}
{Howard}, A.~W., {Bakos}, G.~{\'A}., {Hartman}, J., {et~al.} 2012, \apj, 749,
  134, \dodoi{10.1088/0004-637X/749/2/134}

\bibitem[{{Huang} {et~al.}(2016){Huang}, {Wu}, \&
  {Triaud}}]{2016ApJ...825...98H}
{Huang}, C., {Wu}, Y., \& {Triaud}, A.~H.~M.~J. 2016, \apj, 825, 98,
  \dodoi{10.3847/0004-637X/825/2/98}

\bibitem[{{Jehin} {et~al.}(2011){Jehin}, {Gillon}, {Queloz}, {Magain},
  {Manfroid}, {Chantry}, {Lendl}, {Hutsem{\'e}kers}, \&
  {Udry}}]{2011Msngr.145....2J}
{Jehin}, E., {Gillon}, M., {Queloz}, D., {et~al.} 2011, The Messenger, 145, 2

\bibitem[{{Lendl} {et~al.}(2012){Lendl}, {Anderson}, {Collier-Cameron},
  {Doyle}, {Gillon}, {Hellier}, {Jehin}, {Lister}, {Maxted}, {Pepe},
  {Pollacco}, {Queloz}, {Smalley}, {S{\'e}gransan}, {Smith}, {Triaud}, {Udry},
  {West}, \& {Wheatley}}]{2012A&A...544A..72L}
{Lendl}, M., {Anderson}, D.~R., {Collier-Cameron}, A., {et~al.} 2012, \aap,
  544, A72, \dodoi{10.1051/0004-6361/201219585}

\bibitem[{{Lin} {et~al.}(1996){Lin}, {Bodenheimer}, \&
  {Richardson}}]{1996Natur.380..606L}
{Lin}, D.~N.~C., {Bodenheimer}, P., \& {Richardson}, D.~C. 1996, \nat, 380,
  606, \dodoi{10.1038/380606a0}

\bibitem[{{Maxted} {et~al.}(2015){Maxted}, {Serenelli}, \&
  {Southworth}}]{2015A&A...575A..36M}
{Maxted}, P.~F.~L., {Serenelli}, A.~M., \& {Southworth}, J. 2015, \aap, 575,
  A36, \dodoi{10.1051/0004-6361/201425331}

\bibitem[{{Maxted} {et~al.}(2011){Maxted}, {Anderson}, {Collier Cameron},
  {Hellier}, {Queloz}, {Smalley}, {Street}, {Triaud}, {West}, {Gillon},
  {Lister}, {Pepe}, {Pollacco}, {S{\'e}gransan}, {Smith}, \&
  {Udry}}]{2011PASP..123..547M}
{Maxted}, P.~F.~L., {Anderson}, D.~R., {Collier Cameron}, A., {et~al.} 2011,
  \pasp, 123, 547, \dodoi{10.1086/660007}

\bibitem[{{Neveu-VanMalle} {et~al.}(2016){Neveu-VanMalle}, {Queloz},
  {Anderson}, {Brown}, {Collier Cameron}, {Delrez}, {D{\'{\i}}az}, {Gillon},
  {Hellier}, {Jehin}, {Lister}, {Pepe}, {Rojo}, {S{\'e}gransan}, {Triaud},
  {Turner}, \& {Udry}}]{2016A&A...586A..93N}
{Neveu-VanMalle}, M., {Queloz}, D., {Anderson}, D.~R., {et~al.} 2016, \aap,
  586, A93, \dodoi{10.1051/0004-6361/201526965}

\bibitem[{{Pepe} {et~al.}(2002){Pepe}, {Mayor}, {Rupprecht}, {Avila},
  {Ballester}, {Beckers}, {Benz}, {Bertaux}, {Bouchy}, {Buzzoni}, {Cavadore},
  {Deiries}, {Dekker}, {Delabre}, {D'Odorico}, {Eckert}, {Fischer}, {Fleury},
  {George}, {Gilliotte}, {Gojak}, {Guzman}, {Koch}, {Kohler}, {Kotzlowski},
  {Lacroix}, {Le Merrer}, {Lizon}, {Lo Curto}, {Longinotti}, {Megevand},
  {Pasquini}, {Petitpas}, {Pichard}, {Queloz}, {Reyes}, {Richaud}, {Sivan},
  {Sosnowska}, {Soto}, {Udry}, {Ureta}, {van Kesteren}, {Weber}, {Weilenmann},
  {Wicenec}, {Wieland}, {Christensen-Dalsgaard}, {Dravins}, {Hatzes},
  {K{\"u}rster}, {Paresce}, \& {Penny}}]{2002Msngr.110....9P}
{Pepe}, F., {Mayor}, M., {Rupprecht}, G., {et~al.} 2002, The Messenger, 110, 9

\bibitem[{{Pepper} {et~al.}(2007){Pepper}, {Pogge}, {DePoy}, {Marshall},
  {Stanek}, {Stutz}, {Poindexter}, {Siverd}, {O'Brien}, {Trueblood}, \&
  {Trueblood}}]{2007PASP..119..923P}
{Pepper}, J., {Pogge}, R.~W., {DePoy}, D.~L., {et~al.} 2007, \pasp, 119, 923,
  \dodoi{10.1086/521836}

\bibitem[{{Petrovich} \& {Tremaine}(2016)}]{2016ApJ...829..132P}
{Petrovich}, C., \& {Tremaine}, S. 2016, \apj, 829, 132,
  \dodoi{10.3847/0004-637X/829/2/132}

\bibitem[{{Pollacco} {et~al.}(2006){Pollacco}, {Skillen}, {Cameron},
  {Christian}, {Hellier}, {Irwin}, {Lister}, {Street}, {West}, {Anderson},
  {Clarkson}, {Deeg}, {Enoch}, {Evans}, {Fitzsimmons}, {Haswell}, {Hodgkin},
  {Horne}, {Kane}, {Keenan}, {Maxted}, {Norton}, {Osborne}, {Parley}, {Ryans},
  {Smalley}, {Wheatley}, \& {Wilson}}]{2006PASP..118.1407P}
{Pollacco}, D.~L., {Skillen}, I., {Cameron}, A.~C., {et~al.} 2006, \pasp, 118,
  1407, \dodoi{10.1086/508556}

\bibitem[{{Queloz} {et~al.}(2000){Queloz}, {Mayor}, {Weber}, {Bl{\'e}cha},
  {Burnet}, {Confino}, {Naef}, {Pepe}, {Santos}, \&
  {Udry}}]{2000A&A...354...99Q}
{Queloz}, D., {Mayor}, M., {Weber}, L., {et~al.} 2000, \aap, 354, 99

\bibitem[{{Queloz} {et~al.}(2001){Queloz}, {Henry}, {Sivan}, {Baliunas},
  {Beuzit}, {Donahue}, {Mayor}, {Naef}, {Perrier}, \&
  {Udry}}]{2001A&A...379..279Q}
{Queloz}, D., {Henry}, G.~W., {Sivan}, J.~P., {et~al.} 2001, \aap, 379, 279,
  \dodoi{10.1051/0004-6361:20011308}

\bibitem[{{Ricker} {et~al.}(2014){Ricker}, {Winn}, {Vanderspek}, {Latham},
  {Bakos}, {Bean}, {Berta-Thompson}, {Brown}, {Buchhave}, {Butler}, {Butler},
  {Chaplin}, {Charbonneau}, {Christensen-Dalsgaard}, {Clampin}, {Deming},
  {Doty}, {De Lee}, {Dressing}, {Dunham}, {Endl}, {Fressin}, {Ge}, {Henning},
  {Holman}, {Howard}, {Ida}, {Jenkins}, {Jernigan}, {Johnson}, {Kaltenegger},
  {Kawai}, {Kjeldsen}, {Laughlin}, {Levine}, {Lin}, {Lissauer}, {MacQueen},
  {Marcy}, {McCullough}, {Morton}, {Narita}, {Paegert}, {Palle}, {Pepe},
  {Pepper}, {Quirrenbach}, {Rinehart}, {Sasselov}, {Sato}, {Seager},
  {Sozzetti}, {Stassun}, {Sullivan}, {Szentgyorgyi}, {Torres}, {Udry}, \&
  {Villasenor}}]{2014SPIE.9143E..20R}
{Ricker}, G.~R., {Winn}, J.~N., {Vanderspek}, R., {et~al.} 2014, in \procspie,
  Vol. 9143, Space Telescopes and Instrumentation 2014: Optical, Infrared, and
  Millimeter Wave, 914320

\bibitem[{{Santerne} {et~al.}(2016){Santerne}, {Moutou}, {Tsantaki}, {Bouchy},
  {H{\'e}brard}, {Adibekyan}, {Almenara}, {Amard}, {Barros}, {Boisse},
  {Bonomo}, {Bruno}, {Courcol}, {Deleuil}, {Demangeon}, {D{\'{\i}}az},
  {Guillot}, {Havel}, {Montagnier}, {Rajpurohit}, {Rey}, \&
  {Santos}}]{2016A&A...587A..64S}
{Santerne}, A., {Moutou}, C., {Tsantaki}, M., {et~al.} 2016, \aap, 587, A64,
  \dodoi{10.1051/0004-6361/201527329}

\bibitem[{{Southworth}(2011)}]{2011MNRAS.417.2166S}
{Southworth}, J. 2011, \mnras, 417, 2166,
  \dodoi{10.1111/j.1365-2966.2011.19399.x}

\bibitem[{{Triaud} {et~al.}(2017){Triaud}, {Neveu-VanMalle}, {Lendl},
  {Anderson}, {Collier Cameron}, {Delrez}, {Doyle}, {Gillon}, {Hellier},
  {Jehin}, {Maxted}, {S{\'e}gransan}, {Smalley}, {Queloz}, {Pollacco},
  {Southworth}, {Tregloan-Reed}, {Udry}, \& {West}}]{2017MNRAS.467.1714T}
{Triaud}, A.~H.~M.~J., {Neveu-VanMalle}, M., {Lendl}, M., {et~al.} 2017,
  \mnras, 467, 1714, \dodoi{10.1093/mnras/stx154}

\bibitem[{{Wheatley} {et~al.}(2018){Wheatley}, {West}, {Goad}, {Jenkins},
  {Pollacco}, {Queloz}, {Rauer}, {Udry}, {Watson}, {Chazelas}, {Eigm{\"u}ller},
  {Lambert}, {Genolet}, {McCormac}, {Walker}, {Armstrong}, {Bayliss}, {Bento},
  {Bouchy}, {Burleigh}, {Cabrera}, {Casewell}, {Chaushev}, {Chote},
  {Csizmadia}, {Erikson}, {Faedi}, {Foxell}, {G{\"a}nsicke}, {Gillen},
  {Grange}, {G{\"u}nther}, {Hodgkin}, {Jackman}, {Jord{\'a}n}, {Louden},
  {Metrailler}, {Moyano}, {Nielsen}, {Osborn}, {Poppenhaeger}, {Raddi},
  {Raynard}, {Smith}, {Soto}, \& {Titz-Weider}}]{2018MNRAS.475.4476W}
{Wheatley}, P.~J., {West}, R.~G., {Goad}, M.~R., {et~al.} 2018, \mnras, 475,
  4476, \dodoi{10.1093/mnras/stx2836}

\bibitem[{{Yao} {et~al.}(2018){Yao}, {Pepper}, {Gaudi}, {Beatty}, {Colon},
  {James}, {Kuhn}, {Labadie-Bartz}, {Lund}, {Rodriguez}, {Siverd}, {Stassun},
  {Stevens}, {Villanueva}, \& {Bayliss}}]{2018arXiv180711922Y}
{Yao}, X., {Pepper}, J., {Gaudi}, B.~S., {et~al.} 2018, ArXiv e-prints.
\newblock \doarXiv{1807.11922}

\end{thebibliography}

\end{document}